\documentclass{iaus}
\usepackage{graphicx}

\title[The Tautenburg Exoplanet Search Telescope] 
{TEST \\ The Tautenburg Exoplanet Search Telescope}

\author[Philipp Eigm{\"u}ller \& Jochen Eisl{\"o}ffel]   
{Philipp  Eigm{\"u}ller$^1$
 \and Jochen Eisl{\"o}ffel$^2$}

\affiliation{Th{\"u}ringer Landessternwarte Tautenburg, \\ Sternwarte 5,
DE-07778, Tautenburg, Germany \\[\affilskip] $^1$email: {\tt philipp@tls-tautenburg.de} \\[\affilskip]
$^2$email: {\tt jochen@tls-tautenburg.de}} 

\pubyear{2008}
\volume{253}  
\pagerange{1-4}
\jname{Transiting Planets}
\editors{} 
\begin{document}

\maketitle

\begin{abstract}
The Tautenburg Exoplanet Search Telescope (TEST) is a robotic telescope system. The telescope uses a folded Schmidt Camera with a 300mm main mirror. The focal length is 940mm and it gives a $2.2^{\circ}$ x $2.2^{\circ}$ field of view. Dome, mount, and CCD cameras are controlled by a software bundle made by Software Bisque. The automation of the telescope includes selection  of the night observing program from a given framework, taking darks and skyflats, field identification, guiding, data taking, and archiving. For the search for transiting exoplanets and variable stars an automated psf photometry based on IRAF and a lightcurve analysis based on ESO-Midas are conducted. The images and the results are managed using a PostgreSQL database.
\end{abstract}

\firstsection 

\begin{figure}
\begin{center}
 \includegraphics[width=4.9in]{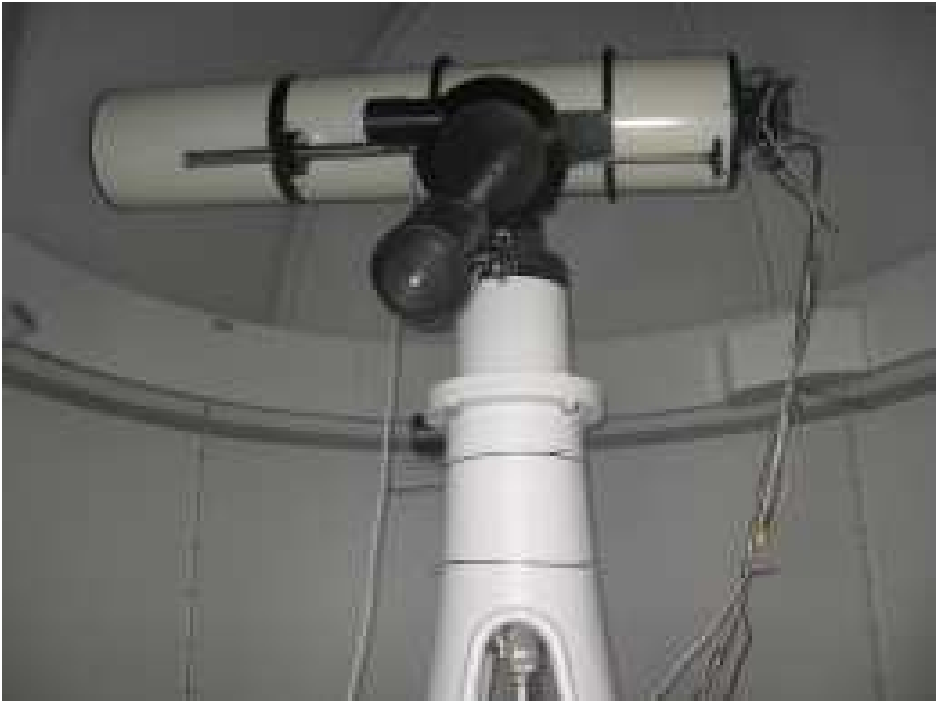} 
 \caption{The Tautenburg Exoplanet Search Telescope (TEST).}
   \label{test}
\end{center}
\end{figure}

\section{Introduction}

In 1999 the first transiting extrasolar planet was discovered (\cite[Charbonneau et al. 2000]{Charbonneau_etal00}). As the knowledge of transit parameters allows an unique insight into the nature of the planet, many efforts have been made to increase the number of known transiting planets. Until now about 50 transiting extrasolar planets are discovered. Inspired by the Berlin Exoplanet Search Telescope (BEST) (\cite[Rauer et al. 2004]{Rauer_etal04}), which was operated by the Deutsches Zentrum f{\"u}r Luft- und Raumfahrt at the Th{\"u}ringer Landessternwarte Tautenburg in the years 2001~-~2003, a new small aperture telescope has now been installed at this site to search for transiting exoplanets.

\section{Instrument}

The Tautenburg Exoplanet Search Telescope, or TEST is a Flatfield camera with 300~mm aperture and 940~mm focal length (see fig. \ref{test}).
It is equipped with an APOGEE CCD camera with 4kx4k pixels with a scale of 2.0 arcsec/pixel, giving a field of view of 2.2x2.2 deg$^2$. The observations are done through an R-band filter. For the guiding a seperate camera is used. The TEST is mounted on a Lichtenknecker M145 mount which is placed in a 3m dome. 

The CCD cameras, mount, and dome are controlled by a software bundle made by Software Bisque. These programms are controlled by scripts which allows complete robotic observations. The weather conditions (humidity, temperature, dewpoint) are also monitored.

\section{Data Analysis}

In the searches of transiting exoplanets huge amounts of high precision photometric data are obtained. For the reduction, photometry, and lightcurve analysis of these data a program written in Python is used. The graphical user interface uses the wxPython framework. 
To allow fast and easy access to the original data and the resulting lightcurves, these are managed using a PostgreSQL database.

The standard reduction is based on the ccdred package of Iraf/Pyraf(\cite[Doug Tody 1986]{Tod86}). The psf-photometry is using the Source Extractor (\cite[Bertini \& Arnouts 1996]{Ber96})to identify stars and the daophot package (\cite[Stetson 1987]{Stel1987}) to determine the brightness of the light sources. The astrometry is done by WCS-Tools using the USNO-A2 catalog.
The algorithms used to analyse the lightcurves are mainly using the ESO-Midas python interface and the SciPy library. The lightcurve analysis consists of five steps: trend filtering, search for general variability, longterm trend identification, search for sinusoidal signals, and the search for transit-like events.

Three trend filters are integrated in the program. SysRem (\cite[Tamuz et al. 2005]{Tam2005}), TFA (\cite[Kov\'{a}cs et al. 2005]{Kov2005}) and an algorithm proposed by \cite[Scholz \& Eisl{\"o}ffel (2004)]{Sch2004}. To get rid of outliers a 3$\sigma$ clipping can be applied to the lightcurves. A combination of these algorithms in any order is possible.
The search for general variability uses the Stetson variability index in its modified version (\cite[Stetson 1996]{Ste1996}, \cite[Zhang et al. 2003]{Zha2003}). In addition, the standard deviation within each lightcurve as a function of the object brightness can be used to derive a second variability index.
Sinusoidal signals are identified by a combination of the Scargle-Lomb periodogram (\cite[Scargle 1982]{Sca1982}) and the CLEAN algorithm (\cite[Roberts et al. 1987]{Rob1987}). Identified binaries can be analysed using the DEBIL/MECI programs (\cite[Devor 2005 ]{Dev2005} and \cite[Devor \& Charbonneau 2006]{Dev2006}). Transit-like events  can be found using the BLS (\cite[Kov\'{a}cs et al. 2002]{Kov2002}) and Matched Filter algorithms (For more details on the lightcurve analysis see \cite[Eig{\"u}ller 2006]{Eig06} and \cite[Eisl{\"o}ffel et al. 2007]{Eis07}.

\begin{figure}
\begin{center}
 \includegraphics[width=4.9in]{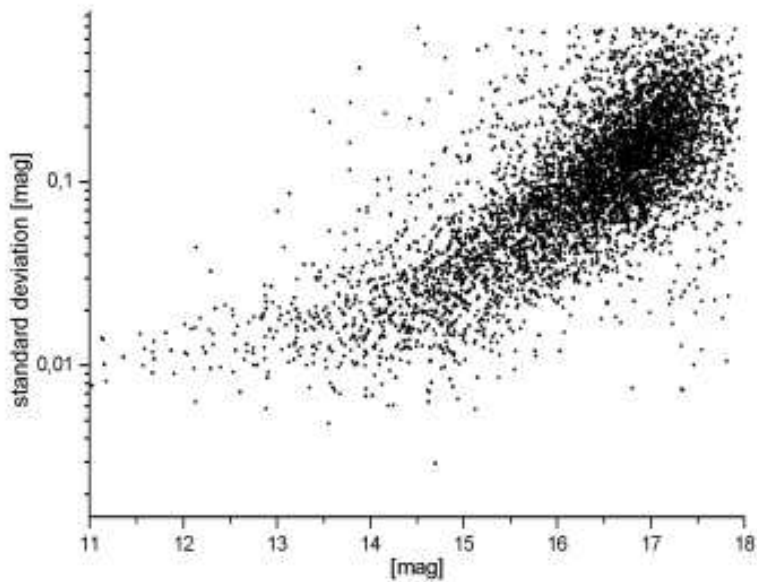} 
 \caption{The standard deviation over the mean magnitude from an early dataset (exposure 120s).}
   \label{rms_mag}
\end{center}
\end{figure}

\section{Observations}
The TEST has started its science programme and is regularly collecting data. This concentrates on a joint transit search with the BEST and BEST II telescopes, and on photometric follow-up for the CoRoT space mission. The TEST achieves the necessary accuracy to detect Jupiter sized planets around Sun like stars in a magnitude range of Vmag = 10-15 magnitudes depending on the exposure time (see fig. \ref{rms_mag}).

\section{Acknowledgements}
The transit work of PE and JE is partially supported by Deutsche Forschungsgemeinschaft grant Ei 409/14-1.\\

PE acknowledges travel support from the IAU to come to this symposium.\\

The used packages Pyraf and Pyfits are products of the Space Telescope Science Institute, which is operated by AURA for NASA.

\end{document}